\documentclass[prl,twocolumn,superscriptaddress,showpacs]{revtex4}

\usepackage{amsmath}
\usepackage{amssymb}
\usepackage{graphicx}
\usepackage[noxcolor]{pstricks}

\newcommand{\CC}{\mathbb{C}}

\begin{document}

\title{Topological entanglement entropy relations for multi phase systems with interfaces}

\author{F.~A.~Bais}
\email{F.A.Bais@uva.nl}
\affiliation{Institute for Theoretical Physics, University of Amsterdam, Valckenierstraat 65, 1018 XE Amsterdam, The Netherlands}
\affiliation{Santa Fe Institute, Santa Fe, NM 87501, USA}

\author{J.~K.~Slingerland}
\email{joost@thphys.nuim.ie}
\affiliation{Dublin Institute for Advanced Studies, School of Theoretical Physics, 10 Burlington Rd, Dublin, Ireland}
\affiliation{Department of Mathematical Physics, National University of Ireland, Maynooth, Ireland}


\begin{abstract}
We study the change in topological entanglement entropy that occurs when a two-dimensional system in a topologically ordered phase undergoes a transition to another such phase due to the formation of a Bose condensate. We also consider the topological entanglement entropy of systems with domains in different topological phases, and of phase boundaries between these domains. We calculate the topological entropy of these interfaces and derive two fundamental relations between the interface topological entropy and the bulk topological entropies on both sides of the interface. 
\end{abstract}

\date{\today}


\maketitle


In gapped two dimensional systems, the ground state entanglement entropy $S_{E}$ of a spatial domain with boundary of length $L$ scales with $L$ as $S_{E} = \alpha L - \gamma + \cdots$. Here $\alpha$ is a constant and the dots represent contributions that vanish in the limit $L \rightarrow \infty$. The additive constant $-\gamma$ is called the {\em topological entanglement entropy} \cite{kitaev-2006-96,levin-2006-96}. It is independent of the detailed geometry of the spatial region (depending only on its topology) and its value is robust against perturbations of the system's Hamiltonian, changing only in phase transitions. In fact, one may express $\gamma$ explicitly in terms of the quantum dimensions $d_a$ of the various types of topological excitations that may occur in the system 
For a region with the topology of a disk, $\gamma = \log D$, with $D= \sqrt{\sum_a d_a^2}$. Topologically trivial phases
have just one topological sector with quantum dimension equal to $1$ and hence have $D=1$ and $\gamma=0$, while all topologically nontrivial phases have $\gamma>0$. Hence 
$\gamma$ has become a popular indicator of the presence of topological order. In fact the size of $\gamma$ is an indicator for the complexity of the topological order or 
topological symmetry present in a system.
In particular, $D$ and the number of topological sectors together
determine whether a system hosts non-Abelian anyons, since such anyons have quantum dimensions greater than $1$.

It is clearly of interest to know how $\gamma$ changes in phase transitions 
In Refs.~\cite{BSS02,BSS03,Bais2008-08}, we have developed a method for the description of topological symmetry breaking transitions between topological phases, caused by condensation of a bosonic excitation. In such transitions, the value of $D$ always decreases. 
Our first aim here is to calculate the corresponding increase in bulk topological entropy in terms of a number $q$ characterizing the transition. After that, we obtain a formula for the topological entropy of a spatial interface between two phases related by condensation. Finally we consider multilayer systems.


Let $A$ and $U$ denote topological phases of the same system, where $U$ is obtained from $A$ by topological symmetry breaking through a Bose condensate. We will abuse notation and call the quantum groups, unitary braided tensor categories or fusion algebras of these phases $A$ and $U$ as well, as long as no confusion can be expected. 

Since $\gamma_A=\log D_A$ and $\gamma_{U}=\log D_{U}$, the topological entropy increases by
\begin{eqnarray}
\gamma_{A}-\gamma_{U}= \log \frac{D_A}{D_{U}}.
\end{eqnarray}
We will now derive an expression for this number in terms of data directly related to the condensation transition.

The features of the description of topological symmetry breaking are as follows (see Refs.~\cite{BSS02,BSS03,Bais2008-08}). 
We assume that a bosonic excitation of phase $A$ condenses. The new, condensed, topological phase has particle like (localized) excitations whose fusion and braiding is described by quantum group $U$.
However, the condensed phase also allows  for confined excitations which pull physical strings in the condensate. In a system with domains of both phases, the confined excitations are expelled to the boundary of the condensed domain. On this boundary there is a well defined set of fusion rules for both confined and unconfined excitations, described by a fusion algebra $T$. While $T$ allows for consistent (associative) fusion, it will not usually allow for a consistent implementation of the braid group, exactly because the confined excitations are attached to the boundary by strings. Our scheme therefore involves three levels, or two steps. We start with the unbroken phase described by the quantum group $A$, which by the condensate gets broken to an algebra $T$. A subset of the $T$-sectors corresponds to unconfined excitations. These are described by a quantum group $U$

The breaking step is described by a branching of topological charge sectors of the $A$-phase into sectors of the intermediate fusion algebra $T$, 
\begin{equation}
a \to \sum_t n^t_a t ~~~~~~ \mathrm{(restriction)}.
\end{equation} 
Here, the $n^{t}_{a}$ are integer coefficients. We can think of this decomposition of $A$-sectors as resulting from \emph{restriction} of representations of the quantum group $A$
to a subalgebra which is not broken by the phase transition.  
Conversely the \emph{lift} of the fields from $T$ to $A$ is defined by
\begin{equation}
	t\to \sum_a n^a_t a ~~~~~~ \mathrm{(lift)},
\end{equation}
where $n_{t}^{a}=n_{a}^{t}$.
Crucially, restriction must commute with fusion in the $A$ and $T$ theories, and preserve the quantum dimensions of the sectors, so $d_a=\sum_{t}n^t_a d_{t}$.


For any topological sector $t\in T$, we can define a number $q$ related to the transition from $A$ to $U$ by
\begin{equation}
\label{eq:quantumindex}
q= \frac{\sum_a n^a_t d_a}{d_t},
\end{equation}
where the $d_i$ which appear are quantum dimensions in the $A$ and $T$ theories.
In the appendix, we show that $q$ does not depend on the choice of $t$. 
In fact, we prove that $q=D_{A}^{2}/D_{T}^{2}$. In addition, we will soon argue on physical grounds that $D_{T}^{2}=D_{A}D_{U}$ (cf.~Eq.~(\ref{eq:dimensionrelation})), so that we have
\begin{equation}
\label{eq:relDIDII}
q=D_A/D_U.
\end{equation}
These facts about $q$ may also be proved rigorously with the use of advanced mathematics, either in the framework of modular tensor categories~\cite{Kirillov02} or using von Neumann algebras \cite{Bockenhauer02}, but we will only use elementary arguments and physical intuition here. 
Given Eq.~(\ref{eq:relDIDII}), we may conclude that the change in topological entanglement entropy between the phases is simply given by
\begin{equation}
\label{eq:entropydifference}
\gamma_{A}-\gamma_{U} = \log q.
\end{equation}
so $q$ characterizes the phase transition in the sense that it determines the accompanying loss of topological entropy. 

In Ref.~\cite{Bais2008-08}, we argued that, when $A$ and $U$ are topological phases which can be described by conformal field theories (CFTs), quantum group symmetry breaking from $A$ to $U$ can occur if there exists a conformal embedding $\mathcal{A}\subset \mathcal{U}$ of the chiral algebras $\mathcal{U}$ and $\mathcal{A}$ of those CFTs. Therefore, we call $q$ the \textit{quantum embedding index}\footnote{We note that this real valued quantum index is different from the integer Dynkin index $j$ that characterizes conformal embeddings of affine Lie algebras}.  

A simple example of the breaking scheme discussed above occurs for $A=SU(2)_4$, giving $U=SU(3)_1$ upon condensation of the spin $2$, or weight $4$ particle in the $SU(2)_4$ theory. This case correspond to the conformal embedding of the $SU(2)_4$ affine Lie algebra into $SU(3)_1$ and it is extensively discussed in Ref.~\cite{Bais2008-08}. In table~\ref{tab:su2su3T}, we give the sectors of the two theories with their quantum dimensions, as well as those of the intermediate algebra $T$.  One may readily verify that indeed:
\begin{equation}
\label{eq:exampleq }
q=\frac{d_0+d_4}{d_{(00)}}= \frac{d_1+d_3}{d_{1}}=\frac{d_2}{d_{(10)}} =\frac{d_2}{d_{(01)}}=\frac{D_A}{D_U}= 2.
\end{equation}
While $q$ is an integer in this case, usually this will not be true. For example when $A=SU(2)_{10}$ and $U=SO(5)_1$ (see~Ref.\cite{Bais2008-08} for details), one finds that $q=3+\sqrt{3}$.

\begin{table}[!h!!]
\vspace*{4mm}
\begin{center}
$\begin{array}{|c|}
\hline A = SU(2)_4 ~~~~~~~~~~~~~~\Rightarrow~~~~~~~~ T~~~~~~~~~~\Rightarrow U \simeq SU(3)_1\\
{\rm~(unbroken)}~~~~~~~~~~~~~~~~~~~~~~~ {\rm(broken)} ~~~~~~~~~~{\rm ~(after~confinement)}\\
\hline
\hline
\begin{array}{l|ll||l|l||l|l}
0&d_0=1&h_0=0\ & 0\rightarrow 0&d_0=1& (00)& d_{00}=1 \\
1&d_1=\sqrt{3}&h_1=\frac{1}{8}\ & 1\rightarrow 1&d_1=\sqrt{3}& conf. & -\\
2&d_2=2&h_2=\frac{1}{3}&2\rightarrow \left\{ \begin{array}{l }
2_1\\
2_2  
\end{array} \right.&\begin{array}{l }
d_{2_1}=1 \\
d_{2_2}=1  
\end{array} & \begin{array}{l}
(10)    \\
(01)   
\end{array} & 
\begin{array}{l }
   d_{10}   =1    \\
      d_{\bar{01}}=1  
\end{array} \\
3&d_3=\sqrt{3}&h_3=\frac{5}{8}&3\leftrightarrow 1& & &\\
4&d_4=1&h_4=1&4\leftrightarrow 0& & &
\end{array}\\
\hline
\end{array}$
\end{center}
\caption{\footnotesize Topological sectors $\Lambda$ with their quantum dimensions $d_\Lambda$ and topological spins $h_\Lambda$, as they feature in the breaking of an $SU(2)_4$ theory after condensation of $\Lambda=4$ excitations. Decomposition of $A$-sectors into $T$-sectors is indicated. The spins of $U$ particles are the same as for the corresponding $A$-particles. Spins of confined $T$-particles are ill-defined.}
\label{tab:su2su3T}
\end{table}

\begin{figure}[htbp] 
   \centering
   \includegraphics[width=1.5in]{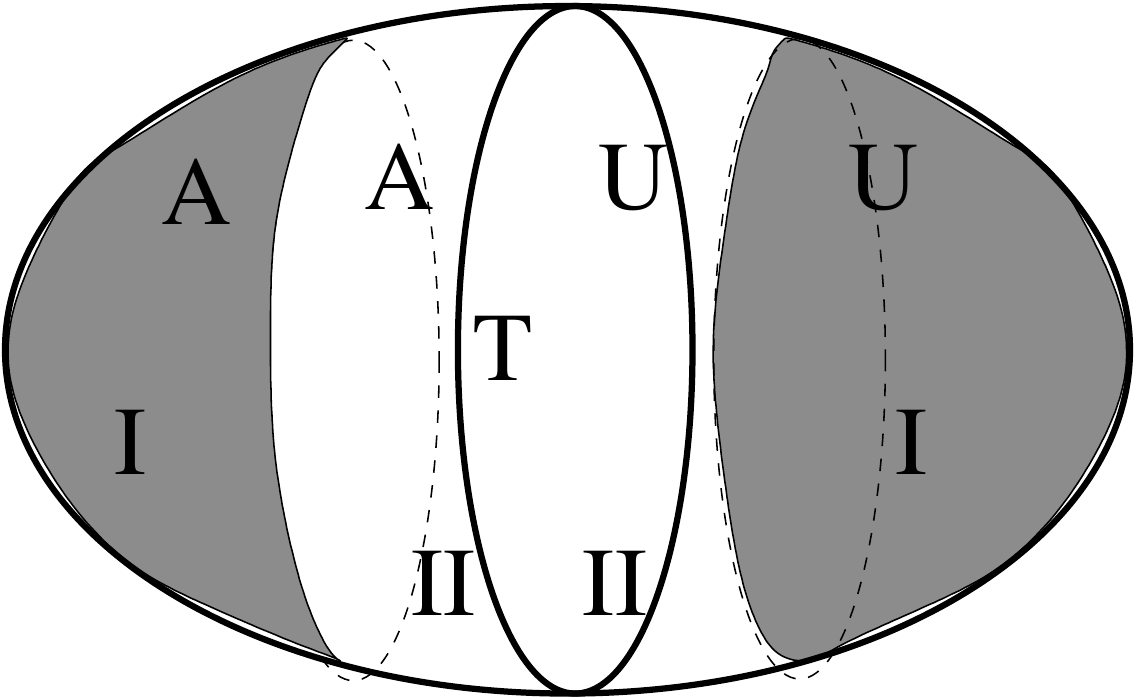} 
   \caption{}
   \label{fig:BreakEntro}
\end{figure}

Let us now consider a geometry as indicated in Fig.~\ref{fig:BreakEntro}, where we have taken a sphere with the left hand side in an unbroken $A$ phase and the right hand side in the condensed $U$ phase, separated by the interface described by $T$. 
We have also divided the sphere into two regions  $I$ (shaded gray) and $II$ for which we wish to calculate the topological entanglement entropy. Since the entanglement entropy of complementary subsystems is equal, we have $S_{E,I}=S_{E,II}$. As the interface geometry is the same for both regions, the topological entanglement entropies must also be equal. Region $I$ consists of two disks, one in the $A$-phase domain and one in the $U$-phase domain, so the total topological entropy here equals $-\gamma_A -\gamma_{U}$. 
We also would like to calculate the entropy of the annular region $II$ directly.  
The topological entropy for an annulus was calculated by Kitaev and Preskill in Ref.~\cite{kitaev-2006-96} and was found to equal twice the entropy for a disk. However, their calculation was done for a system in a single topological phase and does not directly apply here as the region $II$ contains both phases. 
Still, if we narrow the annulus $II$ down to a circle so that only the states on the interface contribute, we would expect to obtain topological entropy $-\gamma_{T}$. These arguments suggest a relation between the bulk and interface topological entropies,
\begin{equation}
\label{eq:entropyrelation}
\gamma_{A} + \gamma_{U}= 2 \gamma_{T}.
\end{equation}
which implies that 
\begin{equation}
\label{eq:dimensionrelation}
D_A D_U=D_{T}^{2}, ~~~\mathrm{or}~~~~D_A/D_T = D_T/D_U.
\end{equation}
We observed this relation already Ref.~\cite{Bais2008-08}, but now we have given it a physical interpretation. {}From Eq.~(\ref{eq:relDIDII}) we also see that $D_A/D_T = D_T/D_U=\sqrt{q}$, so $D$ is reduced by a factor of $\sqrt{q}$ in each of the two steps of the topological symmetry breaking process. 

\begin{figure}[htbp] 
   \centering
    \includegraphics[width=1.5in]{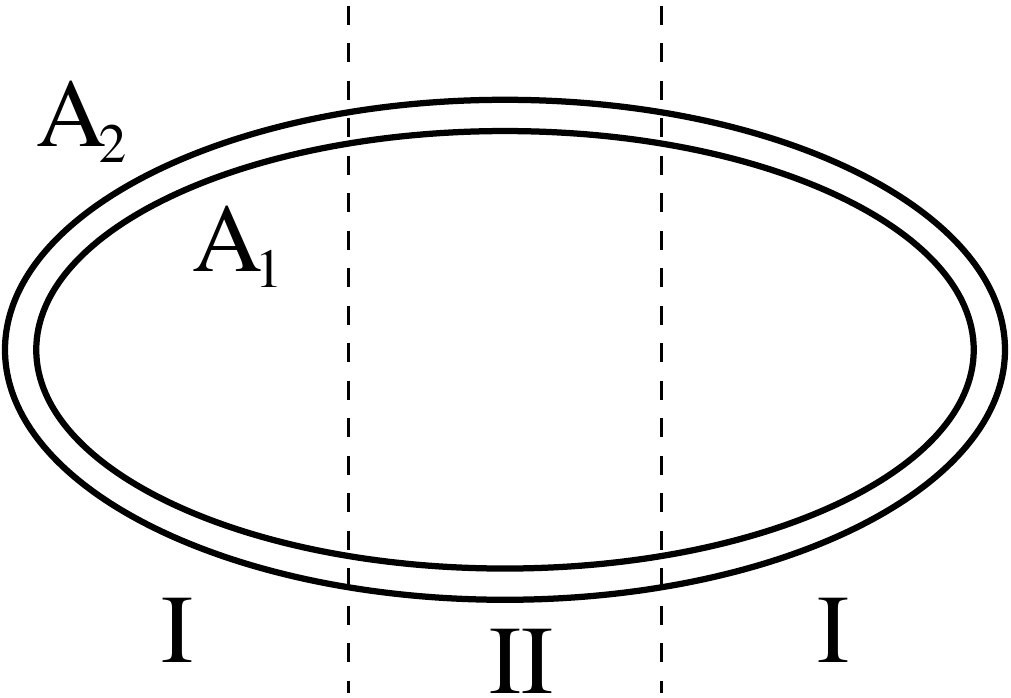}~~~~~ 
   \includegraphics[width=1.5in]{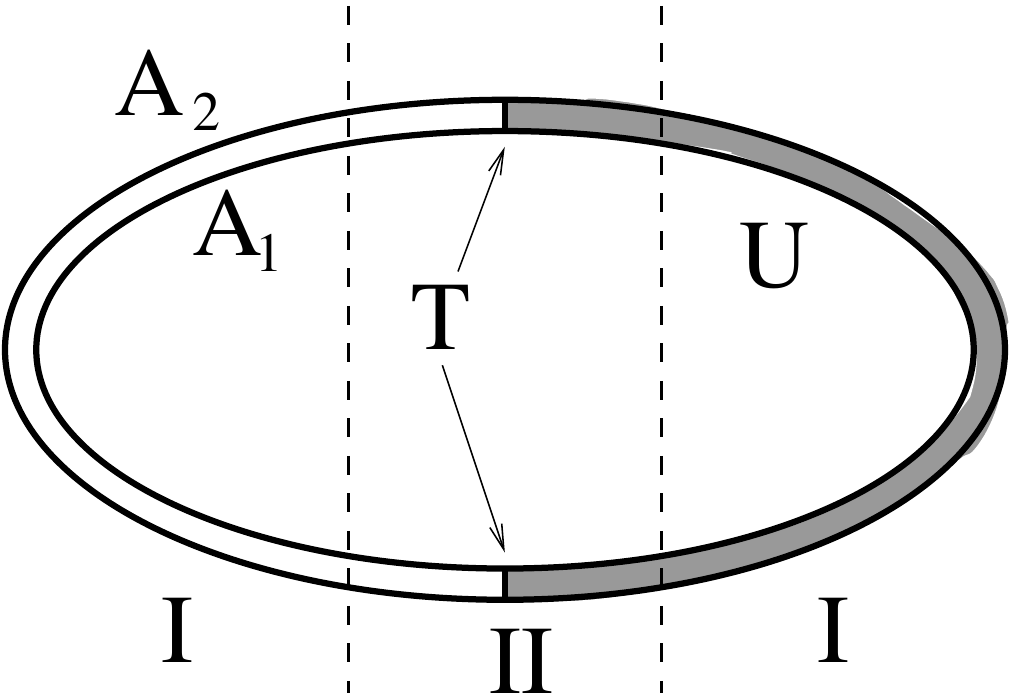}
   \caption{Side view of double layered system with $A\simeq A_{1} \otimes A_{2}$ where on the right a condensate has formed (in gray area). The situation corresponds to the coset construction if $A_1 = G, A_{2}= \bar{H}$ and $U= G/H$.}
   \label{fig:DoubleBreak}
\end{figure}

Now consider a two-layer geometry of two concentric surfaces as indicated in the left picture of Fig.~\ref{fig:DoubleBreak}, with the inner and outer layers in phases $A_1$ and $A_2$, resp.Assume that, in some domain, a Bose condensate forms in a channel that couples the $A_1$ and $A_2$ phases' degrees of freedom. 
We then have a situation as indicated in the right picture of Fig.~\ref{fig:DoubleBreak}, which leads to the entropy relation:
\begin{equation}
\label{eq:DoubleEntropy}
\gamma_{A_1}+\gamma_{A_2} + \gamma_{U} = 2 \gamma_T, ~~~~\mathrm{or}~~~~~D_{A_1} D_{A_2} D_U  = D_T^2.
\end{equation} 
The geometries  depicted in Fig.~\ref{fig:DoubleBreak} can in fact represent the coset construction of CFTs~\cite{Goddard1984}, if we take $A_1 = G, A_{2}= \bar{H}$ and $U= G/H$. 
In that case we have
\begin{equation}
\label{eq:g/hdimensions}
\gamma_G +\gamma_{\bar{H}}+ \gamma_{G/H} = 2 S_T ~~~~\mathrm{or}~~~~~ D_G D_{\bar{H}}D_{G/H} = D_T^2,
\end{equation}
underscoring the role of $T$ in the coset construction.

\begin{figure}[htbp] 
   \centering
   \includegraphics[width=1.5in]{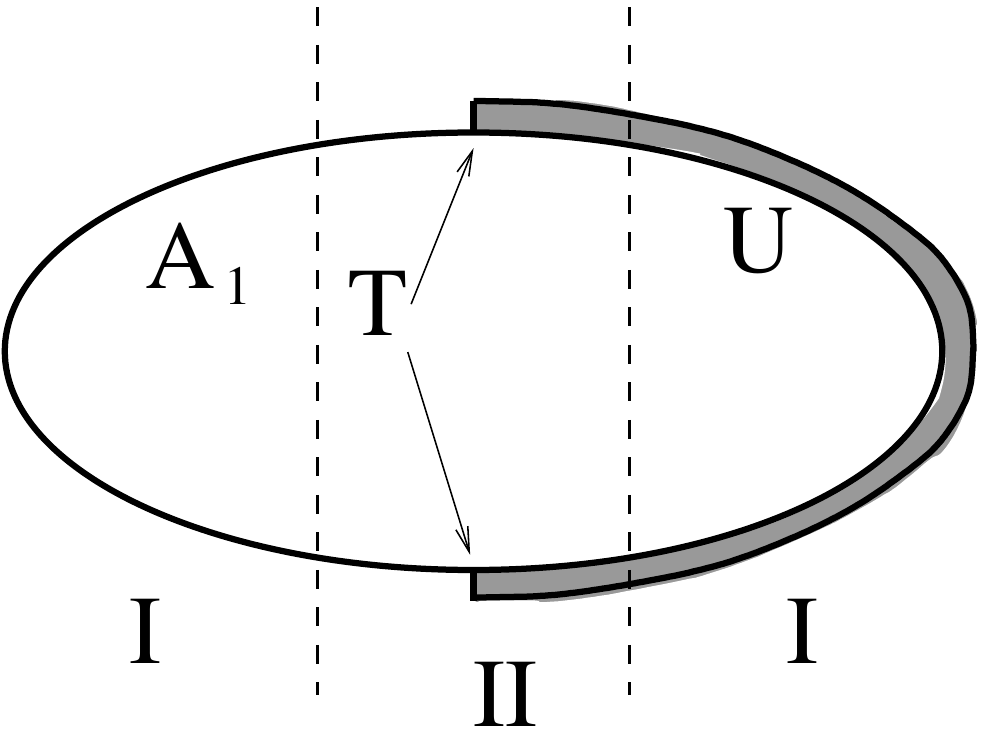} 
   \caption{Single layered system bordering a two layer condensate (gray region). In this situation the central charges of the two domains can be different and gapless degrees of freedom can then exist on the domain boundary.}
   \label{fig:SingleBreak}
\end{figure}

Next imagine that  we take away the disc labeled $A_2$ from the uncondensed area, yielding the situation of Fig.~\ref{fig:SingleBreak}. 
We introduced this setup in Ref.~\cite{BSH08} to describe transitions and interfaces between phases at different central charge $c$, such as those occurring between different fractional quantum Hall conductance plateaus. In Bose condensation transitions, $c$ is conserved, even though $D$ changes. Adding or removing a layer however also adds or subtracts the central charge of that layer. 
One now obtains a modified entropy relation; clearly $\gamma_I$ is lowered by $\gamma_{A_2}$ and therefore the same has to hold for $\gamma_{II}$. Since $T$ has not changed, we find that the interface topological entropy is now $-2\gamma_T + \gamma_{A_2}$. We can heuristically interpret this increase of interface topological entropy as follows. Because of the difference in the central charges on either side, the interface now allows for excitations described by a chiral CFT with the same value of $c$ as the broken layer. In the simplest cases, this is just an $A_2$ type CFT (More detail will appear in~\cite{BBHS09}). The topological entanglement entropy $-\gamma$ normally receives a (negative) contribution from every topological sector of the theory, in gapped bulk phases, where the topological excitations are localized. However the excitations described by the $A_2$ CFT are gapless and delocalized and it appears they do not count toward $\gamma_{II}$.

As an example of a two level system consider $A=A_1\otimes A_2 = {Ising\; \otimes \overline{Ising}}$ and assume a condensate forms in the bosonic $(\epsilon,\epsilon)$ channel. Table \ref{tab:Ising} lists the fields of the intermediate $T$ algebra with their quantum dimensions, and the lifts of the fields into $A$. The fields $(\sigma,1)$ and $(1,\sigma)$ are confined and the remaining fields have $\mathbb{Z}_2\otimes \mathbb{Z}_2$ fusion rules. We have $D_A=4$, $D_T=2\sqrt{2}$ and $D_U=2$, and $q=D_A/D_U=2$. 
\begin{table}[!h!!]
\begin{center}
$
\begin{array}{|l|l|l|l|l|l}
\hline
~~~~T& ~~~d& ~~~~Lift &~~~ U & ~~~d \\
\hline
\hline
(1,1)&d=1& (1,1)+(\epsilon,\epsilon) &  (1,1) & d=1\\
(\epsilon,1)&d=1& (\epsilon,1)+(1,\epsilon) & (\epsilon,1)&d=1\\
(\sigma,\sigma)_1&d=1&(\sigma,\sigma) &  (\sigma,\sigma)_1 & d=1\\ 
(\sigma,\sigma)_2 &d=1&(\sigma,\sigma) & (\sigma,\sigma)_2 & d=1\\
(\sigma,1)&d=\sqrt{2}& (\sigma,1)+(\sigma,\epsilon) & conf. &-\\
(1,\sigma)& d=\sqrt{2}& (1,\sigma) + (\epsilon,\sigma)&conf. &-\\
\hline
\end{array}
$
\end{center}
\caption{\footnotesize The breaking scheme of the $ A = Ising \otimes \overline{Ising} $ theory, after formation of the $(\epsilon,\epsilon)$ condensate: $ A \Rightarrow T= \mathbb{Z}_2 \otimes Ising \Rightarrow U = \mathbb{Z}_2\otimes \mathbb{Z}_2. $ The total quantum dimensions equal  $D_A=4$, $D_T=2\sqrt{2}$ and $D_U=2$. }
\label{tab:Ising}
\end{table}
Removing one of the $Ising$ layers, as in Fig.~\ref{fig:SingleBreak}, we obtain a boundary between a theory of $Ising$ type and one of the toric code universality class. Such a situation can be realized microscopically in the Kitaev hexagonal models (cf.~\cite{Kitaev06a}). The interface between regions will carry a CFT of Ising type and the value of $\gamma_{II}$ will then be reduced by $\log(D_{Ising})=\log(2)$.
\begin{acknowledgments}
The authors  thank Sebas Eli\"ens for clarifying contributions.   
JKS would like to acknowledge the hospitality and support of the University of Amsterdam.
JKS is supported in part by SFI PI grant 08/IN.1/I1961.
\end{acknowledgments}

\subsection*{Appendix: Relations for $\mathbf{q}$ and $\mathbf{D}$}
We will now derive that the quantum embedding index $q$ as defined in Eq.~(\ref{eq:quantumindex}) is independent of $t$ and satisfies $q=D_{A}^{2}/D_{T}^{2}$. We describe both the $A$ and $T$ theories at the level of fusion algebras. For $A$, we write the fusion rules as $a \times b = N^c_{ab} c$ and define the fusion matrices $N_{a}$ by $(N_a)^c_b = N^c_{ab}$.
For $T$ we use the same notations, but with indices $r,s,t,\ldots$ rather than $a,b,c,\ldots$. The fusion coefficients for $A$ and $T$ satisfy $N_{ab}^{c}=N_{\bar{a}c}^{b}$, and for $A$ we also have $N_{ab}^{c}=N_{ba}^{c}$.

We write the restriction as a map $B:A\rightarrow T$, given by $Ba=\sum_{a}n_{a}^{t}t$. The lift is given by the transpose map $B^{\dagger}:T\rightarrow A$. We require that fusion and restriction commute, so $Ba\times Bb=B(a\times b)$, or more explicitly,
\begin{equation}
\label{EQcrs}
	\sum_{c} N^c_{ab} n^t_c  = \sum_{r,s} n^r_a n^s_b N^t_{rs}. 
\end{equation}
We also have conservation of quantum dimension under the restriction, $\sum_t n^t_a d_t = d_a$, as well as $n_{a}^{t}=n_{\bar{a}}^{\bar{t}}$.

In both $A$ and $T$, the quantum dimensions satisfy the fusion rules, e.g.~for $A$, we have 
\begin{equation}
\label{eq:qdimrel}
d_a d_b = N^c_{ab} d_c.
\end{equation}
As a result, the vector $\omega_{A}$ defined by $\omega_{A} = \sum_{a} d_a a$
is a common Perron-Frobenius eigenvector of all the $N_a$, satisfying the eigenvalue equations $N_a \omega_{A} = d_a \omega_{A}$. Similarly, $\omega_{T} = \sum_{t} d_t t$ is a common eigenvector of the $N_{t}$, despite the fact that these do not necessarily commute.

We need the behavior of the Perron-Frobenius eigenvectors $\omega_{T}$ and $\omega_{A}$ under the restriction $B$ and the lift $B^{\dagger}$. We have that,
	$B^{\dagger} \omega_{T} = \omega_{A}$, since, in components, this is just the conservation of quantum dimension under the restriction.
We will now show that also
\begin{equation}
\label{EQbat}
	B \omega_{A} = q \omega_{T}
\end{equation}
For this, let us define the map $M:T \rightarrow T$ by
\begin{equation}
	M\equiv \sum_a M_a= \sum_{at} n^t_a N_t
\end{equation}
$M$ is the fusion matrix of the restriction of $\sum_{a} a$. The matrix elements of $M$ are all strictly positive, since for any $r,s\in T$ there is some $t\in T$ with $N_{tr}^{s}>0$ and also  any $t\in T$ occurs in the decomposition of some $a\in A$. Therefore the eigenspace of $M$ corresponding to the Perron-Frobenius eigenvalue $\sum_a d_a$ is non-degenerate and spanned by $\omega_{T}$. Hence, if we can show that 
\begin{equation}
\label{eq:Mev}
	M B \omega_{A} = (\sum_a d_a) B \omega_{A}
\end{equation}
then it follows that $B \omega_{A} = \lambda \omega_{T}$ for some $\lambda\in \CC$.
In components, Eq.~(\ref{eq:Mev}) reads
\begin{equation}
	\sum_{a,c,s,r} d_c n^r_a n_s^c N^s_{rt} = \sum_{a,b} d_a d_b n_t^b
\end{equation}
%
To prove this, one can massage the left hand side using $N_{rt}^{s}=N_{\bar{r}s}^{t}$ and $n_{a}^{r}=n_{\bar{a}}^{\bar{r}}$. Then one applies (\ref{EQcrs}), uses that $N_{\bar{a}c}^{b}=N_{ab}^{c}=N_{ba}^{c}$ and finally one applies~(\ref{eq:qdimrel}). Some intermediate results are given below,
\begin{eqnarray*}	
\sum_{a,c,s,r} d_c n^r_a n_s^c N^s_{rt} &=&	
\sum_{a,c,s,r} d_c n_{\bar{a}}^{r}n_{c}^{s}N_{rs}^{t}=
\sum_{a,b,c} d_c N_{\bar{a}c}^{b}n_{b}^{t} \\	
&=&\sum_{a,b,c} d_c N_{ab}^{c}n_{b}^{t}=	
\sum_{a,b} d_a d_b n_t^b.
\end{eqnarray*}
Hence $B \omega_{A} = \lambda \omega_{T}$. Writing this in components, we find
\begin{equation}
	\lambda = \frac{\sum_a n^a_t d_a}{d_t}=q  \nonumber
\end{equation} 
with $q$ as defined in~(\ref{eq:quantumindex}). This shows that $q$ is independent of the choice of $t$. Since $q$ is the quotient of the quantum dimension of a field $t$ in $T$ (or in $U\subset T$) by the quantum dimension of its lift in $A$, we call it the \emph{quantum embedding index}. Another reason for this terminology is the similarity to the embedding index introduced by Dynkin in his work on the classification of  subalgebras of compact Lie algebras. This index satisfies a similar formula, where the quantum dimensions should be replaced by the indices of the Lie algebra representations.

From $B\omega_{A}=\omega_{T}$ and $B^{\dagger}\omega_{T}=q \omega_{A}$, it follows that $B^{\dagger}B$ and $BB^{\dagger}$ satisfy 
\begin{equation}
	B^{\dagger} B \omega_{A} = q \omega_{A} 
	~~~~\mathrm{and}~~~~
	B B^{\dagger}\,\omega_{T}	=q \omega_{T} 
\end{equation}
{}From this we obtain that $q=D_{A}^2/D_{T}^{2}$ as follows,
\begin{equation} 
\label{EQdadt}
	D_A^2 = \left< a^* | a^*\right> = \left< t^*|B^{\dagger}B |t^*\right> = q  \left< t^*|t^*\right> = q D_T^2.
\end{equation}
Combining this relation with~(\ref{eq:dimensionrelation}), one also gets~(\ref{eq:relDIDII}).
%
%
%

\end{document}